# Highly Efficient Memory Failure Prediction using Mcelog-based Data Mining and Machine Learning

Chengdong Yao [0000-0003-2166-7878]

University of Technology, Sydney (UTS), Ultimo, NSW 2007, Australia
`chengdong.yao-1@student.uts.edu.au`

**Abstract.** In the data center, unexpected downtime caused by memory failures can lead to a decline in the stability of the server and even the entire information technology infrastructure, which harms the business. Therefore, whether the memory failure can be accurately predicted in advance has become one of the most important issues to be studied in the data center. However, for the memory failure prediction in the production system, it is necessary to solve technical problems such as huge data noise and extreme imbalance between positive and negative samples, and at the same time ensure the long-term stability of the algorithm. This paper compares and summarizes some commonly used skills and the improvement they can bring. The single model we proposed won the top 14th in the 2nd Alibaba Cloud AIOps Competition belonging to the 25th PAKDD conference. It takes only 30 minutes to pass the online test, while most of the other contestants' solution need more than 3 hours. Codes has been open source to https://www.github.com/ycd2016/acaioc2.

## 1 Dataset

The data used in this paper is the error log of Dynamic Random-Access Memory (DRAM) reported by mcelog provided by Tianchi [1], Alibaba. Among them, the collection time of the training set is from January 1 to May 31, 2019, and the test set is from August 1 to August 10 of the same year. Mcelog [2] is a standard tool for Intel-based Machine Check Architecture (MCA) to record DRAM errors in Linux systems. There are a total of four tables, each row in the table is a sample. The test set has only Table 1-3, and each column has the same meaning as the training set. The test set needs to be obtained using the Application Programming Interface (API) specified by the competition group, and only 1 minute of data can be obtained at a time. To prevent the use of future data, the next minute's data can be obtained only after the competitor's algorithm submits the current prediction results.

## 2 Evaluation metrics

### 2.1 Performance

$$n_{tpp} = \sum_{n=1}^{F} \begin{cases} sigmoid(\frac{pti}{ati}), & pti \leq ati \\ 0, & pti > ati \end{cases}$$

$$sigmoid(x) = \frac{e^x}{e^x+1}$$

$$Precision = \frac{n_{tpp}}{n_{pp}}$$

$$Recall = \frac{n_{tpr}}{n_{pr}}$$

$$score = 100 \times F_{1-score}$$

F1-Score is used as the evaluation function in this competition, and the related terms are defined as follows:

- F: Collection of servers that are predicted to fail and do fail within 7 days.
- pti (predicted time interval): Predict that the server will fail within this interval, in minutes.
- ati (actual time interval): The time when the failure occurs minus the time the failure is predicted, in minutes.
- $n_{pp}$: The number of servers that are predicted to fail within 7 days.
- $n_{tpr}$: The number of servers that are predicted to fail and do fail within 7 days.
- $n_{pr}$: The number of servers that do have memory failures.

Because in the case of pti <= ati, the maximum value that sigmoid can achieve is about 0.731059, so the maximum value that F1-Score is about 0.844641. This means that the minimum score of the competition is 0, while the full score is 84.4641, which is not the usual 100. The higher the score, the better the performance.

### 2.2 Efficiency

The competition requires that the results of every 1-minute log (including multiple servers) must be returned within 5 seconds, including data preprocessing, feature extraction, model reference, and post-processing. Besides, to better measure the efficiency of the algorithm, smps (server minutes per second) is used as a unit of efficiency measurement. The higher the smps value, the more servers or minutes the algorithm can predict per second, that is, the more efficient it is.

As for the hardware of the online test bench, the competition group provides Intel Xeon Platinum 8163 CPU, with up to 4 cores and 8 threads, 32GB memory, and Nvidia V100 GPU, up to 16GB video memory.

## 3 Starter model

### 3.1 Baseline

We designed and implemented such a simple model as the basis for all comparisons. To reduce memory consumption, in this model, features are extracted from three tables, and then merge the samples with the same serial number and time of the server together. Referring to the work of our predecessors [3, 4, 5], the features of this model include four parts:

1. It can be speculated that memory failures may be caused by the accumulation of many minor errors, so we count the number of times different MCA ID and transaction have occurred.
2. Errors in some locations are more likely to lead to failures, while others may have little impact, so the number, times and variance of the locations where errors occur at each level are counted.
3. Considering that the Boolean information in the kernel error log may also be helpful, the true values in these data are also counted.
4. Different brands of memory may have different tolerance to errors, and some brands of memory may be more prone to hardware failures.

The way to divide the data is as follows: exclude the positive sample with an actual time interval greater than 1440 minutes and set the ati of the negative sample to -1 minute. To simplify the requirements, we divide the remaining data into two parts, one is negative samples with ati of -1, and the rest is positive samples. In this way, the original task is transformed into a classification task. As for pti, all output 1 and will predict it in subsequent models. This is because the range of target value is too large, it will be difficult to predict an integer between 1 and 10080 (1440 * 7), and the performance may not even be as good as output 1 on this simple model.

Using the method described above, 5 CatBoost [6] classifiers are trained on the training set using 5-fold cross-validation. In the prediction stage, the mean value of the results of all models is taken as the final result. This baseline scored about 24 points in the online test, and the specific experimental results are listed in section 7 below.

### 3.2 Objective optimization

The above baseline appears to perform well in the training set, but once it is submitted for evaluation, there is a big gap (training-validation gap up to 9.1983). This is because the loss function that the baseline is trying to minimize is Logarithmic Loss, and the performance evaluation function is F1-Score. To reduce this gap and improve the convergence of the model, the whole task is divided into two small tasks and designed a more targeted loss and evaluation function.

First of all, we use the complete data set to train a classifier to divide the server into two categories: will fail and will not fail within a day. The classifier uses Logarithmic Loss as the loss function to be minimized and F1-Score as the evaluation function to maximize (it's just baseline). Then train another regressor with the data that will fail within a day to predict how many minutes the server will work properly. The regressor uses our designed Directed Square Error (DSE) loss function and True Positive Score (TPS) evaluation function, which are described in detail below.

**Directed Square Error:**

$$DSE(p,t) = \begin{cases} (p-t)^2, & p \leq t \\ 10(p-t)^2, & p > t \end{cases}$$

$$DSE'(p,t) = \begin{cases} 2(p-t), & p \leq t \\ 20(p-t), & p > t \end{cases}$$

$$DSE''(p,t) = \begin{cases} 2, & p \leq t \\ 20, & p > t \end{cases}$$

When the predicted value is not greater than the truth, DSE is the square of the difference. The smaller the predicted value, the greater the difference and the greater the DSE. When the predicted value is greater than the truth, DSE is stretched to tenfold of the square of the difference. Because this data point will not get any score in the actual evaluation, the importance is characterized by the asymmetry of the error.

Another reason for choosing this loss function is that CatBoost and LightGBM use the Newton-Raphson method instead of Stochastic Gradient Descent (SGD), which requires the loss function to be second-order derivable [7] (as shown above). The Sigmoid function is also considered, but its second derivative is complex, and the amount of computation is large. The experimental results show that the performance is not as good as DSE.

**True Positive Score:**

$$TPS(p,t) = \sum_{i=1}^{n} \begin{cases} sigmoid(\frac{p_i}{t_i}), & p_i \leq t_i \\ 0, & p_i > t_i \end{cases}$$

TPS is a variant of ntpp described in section 2.1 above. The difference is that ntpp accumulates on the set of servers that are expected to fail and do fail within 7 days, while TPS accumulates on full data. This is because when training this regressor, the samples used are not filtered by the previous classifier, so the set operation in section 2.1 is not applicable.

**Result:**

The result (the online test score dropped to only a little higher than 22) shows that the performance improvement brought by the introduction of such a regressor is none. This indicates that using the features designed above, it is tricky to predict how many minutes the server will work properly. During the training phase, the performance was improved by 1 point. however, the score of the online test dropped by 2 points, making the training-verification error even greater. Just as bad, the efficiency of the models is reduced by about 18%. Therefore, in the following work, we give up this two-stage strategy and only use classifiers.

### 3.3 TabNet

TabNet [8] is a Deep Attention Network proposed as a substitute for Gradient Boosting Decision Tree (GBDT). It can be used to solve the task of classification and regression of tabular data. Because features have been constructed, they can be used directly to test the performance of TabNet on this dataset. Due to the large range of values of our feature (from 0 to $2^{17}$), it does not perform the capabilities that TabNet should have. Moreover, the process of constructing features itself will lead to the loss of information, so the network cannot learn from the complete data.

**Result:**

As expected, TabNet performed poorly on this data set, with a score of less than 22. Whether in the training set or online test, the performance is not as good as the CatBoost one with the same features. Though it is 2.4 points lower in the online test, the speed is even slower, even if it is accelerated by CUDA.

## 4 Data cleansing

### 4.1 Noisy data

The noise in this task mainly comes from two aspects: the uncertainty of the location of memory errors and the uncontrollable factors in the ultra-long prediction time interval.

The memory error location in this dataset is parsed from the log reported by MCA. There are times when an error does occur, but MCA does not know the exact location. Because of the huge memory address space, the average probability will lead to almost zero. So, whenever this happens, the error location will be recorded as -1. This is equivalent to the loss of part of the position information, thus introducing noise. However, it is not easy to deal with this noise. Trying to filter out unknown locations and use only the remaining records resulted in lower performance than before filtering, so that attempt was not submitted for testing. To make the model make better use of this location information, it is necessary to continue the research on this topic in the future.

Another important source of noise is that the time interval is as long as 7 days, so long that other accidents may occur midway and lead to memory failure. Such occasional events are unpredictable, and there are indeed samples in the training data that are the same as the log but with different results of the failure. Our baseline narrows the time interval to 1 day, but this is not necessarily the best value. So, we made a statistical analysis of the ati of the true positive samples detected by the model and found that the maximum value was only about 34 hours, and 75% of the samples were less than 588, which is equivalent to 9 hours and 48 minutes. Therefore, it can be considered that our baseline cannot effectively predict events outside this time interval. Further, the time interval is narrowed down to 9 hours and retrained. After that, a reduced interval model is obtained, and its performance is shown in section 7.

**Result:**

By shortening the time interval, the score of the model is improved by about 4.2 points in the training phase. In the online test, it also improved by about 1.8 points to more than 26, while the speed is almost unchanged. This is reasonable because no new features have been added, the complexity of the model remains the same. This once again shows that the features we constructed in section 3 cannot effectively cope with such a long time interval, and also confirms that accidental events are likely to affect the final result in these 7 days Due to the good performance of doing so, the following work is treated with the same reduced time interval if not specified.

### 4.2    Unbalanced data

The dataset for this task can be judged to be extremely unbalanced. After filtering in section 3.1, the positive sample is less than 50 thousand, but there are more than 2 million negative samples. Such an extreme sample ratio may lead to a large number of false negatives, which makes the logarithmic loss value seem to decline, but there is no substantial improvement in the performance of the model. On the other hand, the data of the training set and verification set should obey the same distribution. It is true that in the actual production environment, as the training set, memory itself is not a fault-prone part. If the sample ratio of the training set is forced to balance, it will lead to a large number of false positives, so it is a dilemma.

After experimental comparison, the results show that the distribution of positive and negative samples to retain the original ratio is better, that is, the original ratio cannot be greatly changed. Because the filtering in section 3.1 resulted in the loss of about 42% of the positive samples, it was necessary to use 50% random sampling on the negative samples and retrained to get the resampled model.

**Result:**

After the ratio of positive and negative samples is adjusted to be roughly the same as the original data, the performance of the model is improved. Especially in the training set, the score of the online test has been improved by about 4.6 points, and the score of the online test has also increased by about 1.1 points, with a result of more than 27. This shows that when dealing with unbalanced data, it is necessary to follow the distribution in the actual production environment. Given this, in the work described below, the rule of keeping the ratio of positive and negative samples roughly the same as the original data is followed. As for the slower speed, because the size of the trained model increases compared with that after reducing the interval and before resampling, we speculate that this is due to the complexity of the internal decision structure of the model.

## 5 Feature engineering

### 5.1 Cross-table and derived features

The work in sections 3.1 to 4.2 above is based on the baseline, that is, it uses the same features as the baseline, but is improved in other ways. However, feature engineering is a very important step for traditional data mining, and our previous work is to prepare an appropriate framework. Only when all the factors except features are fixed, the usefulness of our new features can be effectively judged. In this section, referred to works of many predecessors [3, 4, 5], we made some other useful features based on section 3. They are:

5. Ratio values from different sources of errors are useful. This series can be divided into three categories: the correlation ratio between the number of memory errors checked out by MCA, the number of memory errors that can resolve specific error locations, and the number of hardware errors reported by the Linux kernel log.
6. The time difference between errors can provide information. After all, there is a complete difference between a large number of errors in a short period and a small number of errors in a long time. The error rate and dispersion are expressed by calculating the mean and variance of the time difference between errors.
7. The number of errors that occur in different locations is also useful. Although the variance of the location has been considered to reflect the dispersion of the error location, the variance cannot well reflect the quantity, so adding different numbers of errors at different levels is still valid.
8. The quintuple is itself a piece of information. This quintuple encodes the specific coordinates in which the error occurred, and if there are a large number of different coordinates, it may be related to future failures. So, we also count the number of different quintuples and use them.

Other feature construction methods are also considered, such as Word Vector Embedding and Term Frequency-Inverse Document Frequency (TF-IDF). In this way, the context information can be extracted from the error log sequence, which may also be helpful to determine whether the server is malfunctioning. However, generally speaking, the performance of these features is not as good as the output vector of the Deep Neural Network (DNN), and the construction speed is slow, which may lead to exceeding the time limit. So, these features are not used yet.

**Result:**
The online test results are surprising, despite scoring more than 28 in the online test, the training-verification gap is as high as 20.2 points, the biggest gap so far. Although it has made great progress in the training set by about 6.4 points, it has only improved by about 1.3 points in the online test, much less than offline. Such a large gap means that there is a serious over-fitting phenomenon in the model, which is likely to be caused by some of our newly introduced features, which may vary greatly in the distribution of training set and online test. Considering that this is a serious problem, we have suspended the original plan, and need to find out the features that lead to the serious overfitting of the model.

### 5.2    Features that lead to over-fitting

The new features introduced in the previous section led to serious over-fitting, so they need to be found by dichotomy and removed. In this way, we can get a more generalized model with fewer features, which is helpful to improve the real performance of the model. A mask is applied to the training set. By changing the position of 0 in the mask to control the model not to use those features, and then we retrain the model. After many experiments and comparing the scores of the training set and online test, the serious over-fitting is caused by the following features:

- Ratio of hardware errors to DRAM errors.
- Count of different errors in specific tuples.
- And its accumulation feature on the last day.

**Result:**

After removing the above features, we retrain the model, and the training results show that the performance of the model has been reduced. This degradation is normal because the previous model is seriously over-fitted. So, it was submitted online. It got a performance improvement and faster in the online test, with a score of more than 29. This shows that the last model did have a serious over-fitting, and after removing some features, it has been solved to a certain extent. As the competition is drawing to a close, the rest is model blending. Since it is a very common operation, it will not be described here. The result of this model is shown in the table in section 7 below.

## 6    Feature importance analysis

The feature importance here measures how much the value of the model output (not the final classification result) will change when only the value of this feature changes. When the value of the feature changes to the same degree, the greater the change of the output value, the greater the importance of the feature. The importance of all features is normalized so that the sum is equal to 100% for intuitive comparison. For details of the algorithm, please see the CatBoost official documentation [9] and its implementation, which will not be repeated here. Because the features we have constructed are divided into points 1-8 above, we first analyze the importance of these 8 types of features.

## 6.1 The importance of each type

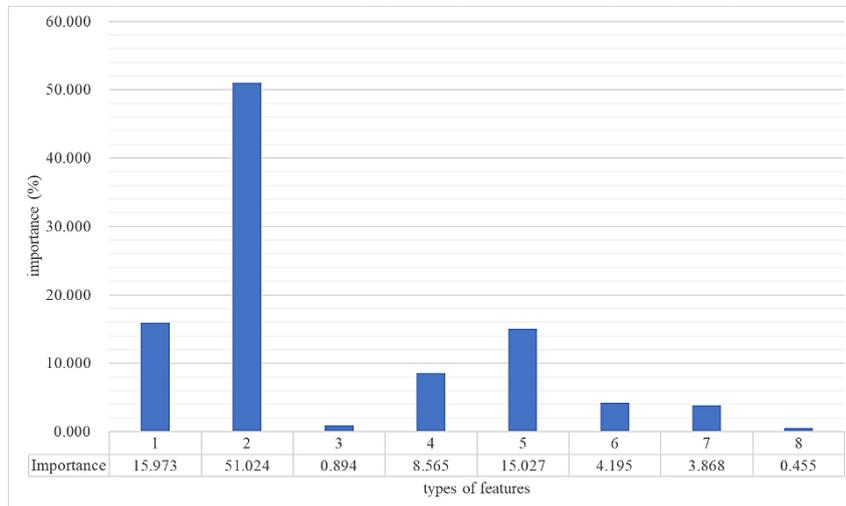

In terms of type alone, the most important one is the type 2 feature, that is, features related to the error location (described in section 3). This shows that to predict whether memory failures will occur, it is important to consider the location, concentration and frequency of memory errors. The type with the lowest importance is the type 8 feature, that is, different quintuples. This may be because the precise error location is of little help in judging whether there will be a failure, or it may be because there are so many location-related features that lead to redundancy.

## 6.2 The 10 most important features

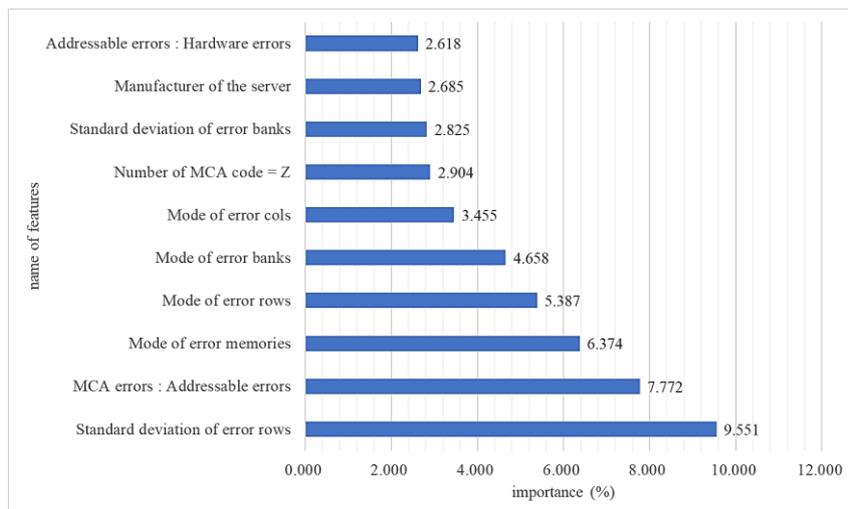

The figure above shows the 10 most important features, more than half of them are location-related features, such as rows, memories, banks, and columns where the error is located. Two of the remaining are ratio features: the ratio of the number of memory errors checked out by MCA to the number of memory errors that can be resolved to a specific location, and the ratio of the number of memory errors that can be resolved to the number of hardware errors reported by the Linux kernel log. The number of errors with MCA code Z after desensitization and the manufacturer of the server (not the vendor of memories) are also important features. These 10 features stand out among a total of 160 features.

### 6.3 Unhelpful features

Some features are of zero importance, which may be due to improper handling, but they do not help the model to predict whether memories will fail at all:

- The number of errors with MCA code AE, AZ or BC after desensitization.
- The difference between the time when MCA or Linux kernel detected the current error and the time when the last error was detected.
- Frequency information about [sel, hwerr_n, hwerr_s, hwerr_m, hwerr_p, hwerr_fl, hwerr_r, hwerr_cd, cmci_sub, hwerr_pi, hwerr_o] related errors reported in the Linux kernel log.

## 7 Comparison of results

**Table. Comparison of results**

| Model | Training performance (Score) | Validation performance (Score) | Efficiency (smps) |
|---|---|---|---|
| **Baseline** | 33.5119 | 24.3136 | 11.7587 |
| **TabNet baseline** | 36.1293 | 21.9653 | 8.5144 |
| **Objectives optimized** | 34.8852 | 22.1432 | 9.6109 |
| **Time interval reduced** | 37.7059 | 26.0961 | 11.7410 |
| **Training set resampled** | 42.3121 | 27.1729 | 10.9224 |
| **Features derived** | 48.6663 | 28.4557 | 9.0758 |
| **Over-fitting solved** | 46.2899 | 29.2701 | 9.4438 |

## Summary

Based on the dataset and evaluation metrics of AIOps competition, we propose a CatBoost-based model to predict whether memory failure will occur in the server for some time in the future. After simple blending, the proposed single model has obtained the ranking of the top 14th and can successfully pass the online test within 40 minutes. In this process, it is very difficult to predict the remaining normal working minutes. Because of the asymmetry of the evaluation measure, the effect of predicting the time interval is not even as good as guessing 1. We designed a series of features based on different aspects, and studied their respective importance, and whether they lead to serious overfitting of the model. The problems of data noise caused by occasional events in a long prediction time interval and the sampling of extremely unbalanced data in the actual production environment are also solved.

However, there are still some problems to be solved. This includes solving the data noise problem caused by the failure of MCA to resolve the address of the specific error and using the context information of the error log to solve the problem of information loss. The latter problem may be solved by using models such as Long Short-Term Memory (LSTM) network or Transformer to extract features from error logs sequences. This topic is worthy of further study to improve the operation and maintenance efficiency of large-scale data centers.

## References

1. PAKDD2021 Alibaba Cloud AIOps Competition, https://tianchi.aliyun.com/competition/entrance/531874/information, last accessed 2021/04/23.
2. Andi Kleen: mcelog: memory error handling in user space. Linux Kongress (2010).
3. Andy A. Hwang, Ioan Stefanovici, Bianca Schroeder: Cosmic rays don't strike twice: understanding the nature of DRAM errors and the implications for system design. ACM SIGARCH Computer Architecture News (2012)
4. Vilas Sridharan, Dean Liberty: A Study of DRAM Failures in the Field. SC '12: Proceedings of the International Conference on High Performance Computing, Networking, Storage and Analysis, pp. 1-11. (2012)
5. Isaac Boixaderas, Darko Zivanovic, Sergi More, Javier Bartolome, David Vicente, Marc Casas, Paul M. Carpenter, Petar Radojkovic, Eduard Ayguade: Cost-Aware Prediction of Uncorrected DRAM Errors in the Field. SC '20: Proceedings of the International Conference for High Performance Computing, Networking, Storage and Analysis, pp. 1-15. (2020)
6. Liudmila Prokhorenkova, Gleb Gusev, Aleksandr Vorobev, Anna Veronika Dorogush, Andrey Gulin: CatBoost: unbiased boosting with categorical features. NeurIPS (2018).
7. Usage examples, https://catboost.ai/docs/concepts/python-usages-examples.html, last accessed 2021/04/23.
8. Sercan O. Arık, Tomas Pfister: TabNet: Attentive Interpretable Tabular Learning. arXiv preprint (2020)
9. Feature importance, https://catboost.ai/docs/concepts/fstr.html, last accessed 2021/04/23.